\documentclass[a4paper,twoside,brazil,english,prd,amsmath,nofootinbib,superscriptaddress,showpacs]{revtex4-1}

\usepackage[T1]{fontenc}
\usepackage[utf8]{inputenc}
\usepackage[brazil,english]{babel}
\usepackage{tensor}
\usepackage[usenames,dvipsnames]{color}
  \definecolor{darkblue}{RGB}{0,0,150}
\usepackage{amssymb}
\usepackage{tensor}
\usepackage{graphicx}
\usepackage{nicefrac}
\usepackage[unicode]{hyperref}
\hypersetup{
  pdfinfo = {
    Author = {A. M. da Silva},
    Title = {Quasilocal approach to general Universal Horizons}
  },
  colorlinks = true,
  breaklinks = true,
 linkcolor = darkblue,
 urlcolor = darkblue,
 citecolor = darkblue
}
\usepackage{multirow}
\usepackage{enumerate}

\newcommand{\ud}{\ensuremath{\mathrm{d}}}
\newcommand{\Lie}{\ensuremath{\mathcal{L}}}

\DeclareMathAlphabet{\mathpzc}{T1}{pzc}{m}{it}

\def\coxa{{\Huge
C\kern-.1667em\lower.5ex\hbox{O}\-X\kern-.1667em\lower.5ex\hbox{A}\@}%
\index{CoXa}
}


\begin{document}

 \title{Quasilocal approach to general universal horizons}
 
\author{Alan Maciel}

\email{alan.silva@ufabc.edu.br}

\affiliation{Centro de Ciências Naturais e Humanas, Universidade Federal do ABC,\\
 Avenida dos Estados 5001, CEP 09210-580, Santo André, São Paulo, Brazil.}

 \begin{abstract}
 Theories of gravity with a preferred foliation usually display arbitrarily fast signal propagation, changing the black hole definition. A new inescapable barrier, the universal horizon, has been defined and many static and spherically symmetric examples have been studied in the literature. Here, we translate the usual definition of the universal horizon in terms of an optical scalar built with the preferred flow defined by the preferred spacetime foliation. The new expression has the advantages of being of quasilocal nature and independent of specific spacetime symmetries in order to be well defined. Therefore, we propose it as a definition for general quasilocal universal horizons. Using the new formalism we show that there are no universal analog of cosmological horizons for FLRW models for any scale factor function, and we also state that quasilocal universal horizons are restricted to trapped regions of the spacetime. Using the evolution equation, we analyze the formation of universal horizons under a truncated Ho\v rava-Lifshitz theory, in spherical symmetry, showing the existence of regions in parameter space where the universal horizon formation cannot be smooth from the center, under some physically reasonable assumptions. We conclude with our view on the next steps for the understanding of black holes in nonrelativistic gravity theories.
 \end{abstract}
 \maketitle
 
 \section{Introduction}
 
 Black holes were originally found and defined as a result of local Lorentz invariance (LLI) in solutions of general relativity (GR). LLI guarantees that the speed of light $c$ is the highest speed available for physical signals what implies, by itself, that the future of any event lies inside a light cone. Black holes appear in solutions of GR that contain a set of events whose future light cones do not reach future infinity and their \emph{event horizon} is  defined as the boundary of the set of events whose future light cones contains future infinity. Therefore, it is surprising that when we build nonrelativistic gravity theories by giving up LLI and allowing for superluminal propagation of physical signals, we can still define black holes and find solutions containing such objects, as was first clearly stated in Refs. \cite{Barausse:2011pu, Blas:2011ni}. The event horizon for black holes in nonrelativistic theories has been called the \emph{universal horizon} (UH), for it is a horizon for signals of arbitrarily large speed, including instantaneous signals. This is possible because even in the absence of a maximum speed, there still is a notion of causality, as it was explained in Ref. \cite{Bhattacharyya:2015gwa}.
 
 On the other hand, gravity theories without LLI have been subject of interest for a variety of reasons, ranging from pure phenomenology applications of Einstein-\ae ther theory \cite{Jacobson:2000xp,Jacobson:2008aj}, alternative models of gravity under different fundamental symmetries from those of GR, such as the Shape Dynamics \cite{Gomes:2010fh} and even as renormalizable quantum gravity theory candidates such as the Ho\v rava-Lifshtiz gravity \cite{Horava:2009uw,Horava:2010zj,daSilva:2010bm,Blas:2009qj}. 
 
Black hole physics is already a challenging subject with deep theoretical consequences in standard GR, such as the relation between black holes thermodynamics \cite{bekenstein-1973, hawking-1976} and Hawking radiation \cite{hawking-1975}, and the current progress on holographic duality between gravity and gauge theory \cite{'tHooft:1993gx,Maldacena:1997zz,Nastase:2007kj}. These important phenomena related to black holes were originally obtained considering only static black hole models, or the phase space of static solutions. In order to study the actual physical evolution of black holes and related physics, the quasilocal formalism based on the optical scalars related to the flow of null curves was developed in the last two decades \cite{Hayward:1993mw, Hayward:1997jp, Hayward:1998hb,Andersson:2007fh, Hayward:2008jq, Senovilla:2011fk, Senovilla:2014ika}.
 
 Recent works have searched for black holes laws in nonrelativistic gravity in analogy with the black hole laws in GR, such as the emission of a Hawking radiation from the UH \cite{Berglund:2012fk} (whose temperature is related to a surface gravity in the standard way) \cite{Cropp:2013zxi,Cropp:2013sea} and the derivation of a Smarr formula \cite{Berglund:2012bu}. Static solutions containing a UH have been studied in Refs. \cite{Lin:2014eaa,Lin:2014ija,Bhattacharyya:2014kta,Ding:2015kba}, while the examples of its dynamical evolution have been analyzed by numerical methods in Refs. \cite{Saravani:2013kva,Tian:2015vha}. While there has been a steady progress in the understanding of UHs, most of the work until now has depended on definitions that are restricted to highly symmetrical setups, such as spherical symmetry, asymptotic flatness, and (with the few exceptions cited above) no time evolution.
 
 Our aim in this paper is to provide a quasilocal formalism that will allow us to study black holes in theories with a preferred foliation, in order to be able to work in nonspherical and time-dependent models, in analogy with the program of quasilocal trapping horizons in relativistic gravity. 
  
 In Sec.~\ref{sec:contexto}, we explain the main assumptions on the spacetime structure that emerge in the class of gravity theories considered in this work. We also define the optical scalars, which are the geometrical objects that we use in the formalism we propose. In Sec.~\ref{sec:static} we review the original definition of a UH, restricted to static and spherically symmetric solutions. In Sec.~\ref{sec:dynamical} we review a generalization used in Ref. \cite{Tian:2015vha} for UH in spherically symmetric, but dynamical spacetimes. Then, we express the earlier definitions in terms of optical scalars that are not dependent on the symmetries of the solution, obtaining as the result our proposal of a general definition of quasilocal UHs. In Sec.~\ref{sec:results}, we study some consequences of the quasilocal definitions in cosmological and gravitational collapse setups. Finally, we state our conclusions and discuss some ideas of further progress in the understanding of black holes in nonrelativistic theories in Sec.~\ref{sec:conclusions}.
 
 In this paper, we adopt the geometrized unit system with $G = c = 1$, the \emph{abstract index notation} as used in Wald's textbook\cite{wald} and the $( - + + + )$ signature.

 \section{Main assumptions and definitions} \label{sec:contexto}
 
 The interest in theories of gravity with a preferred foliation has been considerable and various theories with this property have been proposed. Here we will not focus on any specific Lagrangian, but on properties shared by this class of theories. Specifically, we assume that the three following assumptions hold:
 \begin{enumerate}
  \item \label{def:sigma}The spacetime has a preferred codimension-one spacelike foliation $\Sigma$ as a fundamental structure.
  
  \item \label{def:ordered} The foliation $\Sigma$ is \emph{totally ordered}, such that we can define a real monotonic \emph{time} function $\tau$ that is constant in each single leaf $\Sigma_{\tau}$.

  \item \label{def:S} Each hypersurface $\Sigma_{\tau}$ can be further foliated in compact spacelike 2-surfaces $\mathcal{S}$.
 \end{enumerate}

 For more precise details on the meaning and consequences of assumptions~\ref{def:sigma} and \ref{def:ordered} above, we refer the reader to Ref. \cite{Bhattacharyya:2015gwa} for  precise definitions. Assumption~\ref{def:S} is necessary because we will relate the current UH definitions with geometrical objects built on the 2-surfaces of $\mathcal{S}$.

  A useful  (and physically meaningful) way of describing a codimension-one spacelike foliation is by using a hypersurface-orthogonal normalized timelike 1-form field $u_a$, which is built in terms of the time function as
 \begin{gather}
  u_a =  \frac{1}{\sqrt{-\nabla^a \tau \nabla_a \tau}}  \nabla_a \tau \,\Rightarrow \, u_{[a} \nabla_b u_{c]} = 0 \,.
 \end{gather}

 We will call the vector field $u^a$ the \emph{preferred flow} associated with $\Sigma$.

For the sake of clarity, we can assume that our theory is given by an Einstein-scalar action $S_{total}$ of the form
 \begin{gather}
  S_{total} = S_{EH} + S_{\phi} + S_{\text{matter}} \, ,
 \end{gather}
\noindent
where $S_{EH}$ is the standard Einstein-Hilbert action of GR, $S_{\phi}$ is the non-Lorentz-invariant term given by a scalar $\phi$ field  and its coupling with the spacetime metric $g_{ab}$, and $S_{matter}$ corresponds to the material sources and their couplings to the gravity fields, $g_{ab}$ and $\phi$. If we use a Lagrange multiplier to impose that $\nabla_a \phi$ is timelike everywhere, this theory will have a preferred spacelike foliation (or a special foliation, albeit not preferred, if we follow the definitions in Ref. \cite{Bhattacharyya:2015uxt}) with surfaces orthogonal to $\nabla_a \phi$. Theories that fall into this mold include Ho\v rava-Lifshitz gravity (see Ref. \cite{Blas:2011ni}), Einstein- \ae ther theory restricted to hypersurface-orthogonal flow \cite{Jacobson:2010mx}, k-essence cosmological models \cite{Bonvin:2006vc}, and cuscuton theory \cite{Afshordi:2006ad}.
 
For instance, in Ho\v rava-Lifshitz gravity, the preferred foliation structure is given by a scalar field $\phi(x^a)$ which has been called the \emph{khronon} \cite{Blas:2010hb,Blas:2011ni}. The leaves of the foliation are given by the level 3-surfaces of the khronon field.

 The foliation $\mathcal{S}$ on the $\Sigma_{\tau}$ surfaces induces a codimension-two foliation on the full spacetime, which divides the tangent space at each event into two subspaces, the one tangent to $\mathcal{S}$, which we call $T(\mathcal{S})$ and the one orthogonal to $\mathcal{S}$, which we call $N(\mathcal{S})$. The subspace $N(\mathcal{S})$ contains $u^a$ and since it is timelike, we can always chose a spacelike unit vector $e^a$ in $N(\mathcal{S})$ in order to build a zweibein, satisfying
  \begin{gather}
  u_a e^a = 0 \,, \quad u^a u_a = -1\,, \quad e^a e_a = 1 \, ,
 \end{gather}
 With the spacetime metric $g_{ab}$ we can build the induced metric on the $\mathcal{S}$-surfaces as
 \begin{gather}
  n_{ab}= g_{ab} + u_a u_b - e_a e_b \, ,
 \end{gather}
\noindent
which allow us to define the 2-expansion $\Theta_{(v)}$, 2-shear $\sigma\indices{_{(v)ab}}$, and 2-vorticity $\omega\indices{_{(v)ab}}$ of any orthogonal vector $v^a$, respectively, as the trace, symmetric traceless and antisymmetric components of its derivative on $\mathcal{S}$:
\begin{gather}
 n\indices{_a^c}n\indices{_b^d} \nabla\indices{_c} v\indices{_d} = \frac{1}{2} n\indices{_{ab}}\Theta\indices{_{(v)}} + \sigma\indices{_{(v)ab}} + \omega\indices{_{(v)ab}}\,,
\end{gather}
\noindent
where 
\begin{gather}
 \Theta_{(v)} = n^{ab} \nabla_a v_b = \frac{1}{2}n^{ab} \Lie_v n_{ab} \,, \label{theta-definition}\\ 
 \sigma\indices{_{(v)ab}} = n\indices{_a^c}n\indices{_b^d} \nabla\indices{_{(c}}   v\indices{_{d)}}-\frac{ \Theta_{(v)}}{2} n_{ab}\,,\\
 \omega\indices{_{(v)ab}}= n\indices{_a^c}n\indices{_b^d} \nabla\indices{_{[c}}   v\indices{_{d]}} \, .
\end{gather}
\noindent

 Virtually all examples of UHs {in 3+1 dimensions}\footnote{There are examples of UHs in rotating black holes in the literature, albeit in 2+1 dimensions, as in Refs. \cite{Sotiriou:2014gna,Lin:2016myf}. In 2+1 dimensions, rotating solutions preserve circular symmetry, and hence all formalism for spherically symmetric spacetimes still applies, with the due adaptation for the lower dimensionality.}  studied in the literature are located in spherically symmetric spacetimes, where the foliation $\mathcal{S}$ can be built with round spheres $S^2$. In this case, the 2-shear and 2-vorticity vanish and the 2-expansion related to any vector $v^a \in N(\mathcal{S})$ can be given in terms of the areal radius $r$ as
 \begin{gather}
  \Theta_{(v)} = \frac{2}{r} v^a \nabla_a r \, .
 \end{gather}

 \section{Universal horizons in static spacetimes} \label{sec:static}
 
 The static UHs have been defined in static and spherically symmetric solutions and in this section we restrict ourselves to such spacetimes. Therefore, the vector $u^a$ is orthogonal to the spheres of symmetry, which correspond to the leaves of $\mathcal{S}$. In order to build a spacetime basis, we can choose the spatial vector as 
 \begin{gather}     
 e^a \equiv E_{\perp}^{ab}u_b \,,
 \end{gather}
 \noindent
 where $E_{\perp}^{ab}$ is the Levi-Civita tensor on $N(\mathcal{S})$\footnote{The Levi-Civita tensor on $N(\mathcal{S})$ is given by $E_{\perp}^{ab} = \frac{1}{\sqrt{- \det{g_{\perp \, ab}}}} \epsilon^{ab}$, where $g_{\perp \, ab}$ is the induced metric tensor on $N(\mathcal{S})$.}. 
 
 The vector $e^a$ satisfies
 \begin{gather}
  u_a e^a = 0 \,, \quad e^a e_a = 1 \, ,
 \end{gather}
\noindent
and, in the particular case of flat spacetime where $u^a = (1, 0, 0, 0 )$, we have $e^a = (0, 1, 0 , 0)$ by our construction.
 
  Since we consider the possibility of arbitrarily fast signal propagation, causality in this context is modified from its relativistic form. The light cones are deformed locally into \emph{planes} and the distinction between timelike and spacelike vectors disappears.
 
 What remains is the distinction between past- and future-directed signals: only signals propagating to the future (that is, towards increasing time hypersurfaces) according to the local observer given by the preferred flow $u^a$ are physical.
 
 Covariantly, this causality notion means that for a future-directed signal (or particle) with 4-momentum $p^{a}$ we have
\begin{equation}
 p^{a}u_a < 0 \, . \label{causality}
\end{equation}

 
 The condition \eqref{causality} is the usual condition used in general relativity to define future-directed signals according to some local observer $u^a$ chosen to define the future direction. What is not usual here is that this condition also applies to a spacelike $p^a$. We refer the reader to Ref. \cite{Bhattacharyya:2015gwa} for more details on this notion of causality.


In static spacetimes, there exists a Killing field $\chi^a$ in the exterior region that is timelike outside the Killing horizon and spacelike in the interior region. The UH has been defined in static spacetimes as the tube foliated by surfaces where
\begin{gather}
 \chi^a u_a = 0 \, .\label{staticUH}
\end{gather}
\noindent

 This ensures that all future-directed signals propagate towards smaller radii, since $\chi^a \partial_a r =0$ which implies that $u_a =  W \partial_a r$ at the UH. Equation~\eqref{causality} implies $p^a \partial_a r < 0$ for $W > 0$, which is the case when the future is directed inwards. This defines an interior region where even arbitrarily fast signals are imprisoned, defining a black hole in nonrelativistic theories. The case where the future is directed outwards ($W<0$) does not correspond to a black hole type object and will be treated in Sec.~\ref{sec:results}.

\section{Dynamical cases and general definition} \label{sec:dynamical}


The definition \eqref{staticUH} is restricted to stationary spacetimes as it depends on the existence of a timelike Killing field. Since we expect that black holes are the end products of gravitational collapse, it is necessary to find a definition that does not depend on staticity.

This has been attempted in Ref. \cite {Bhattacharyya:2015gwa}, where a definition of UHs independent of staticity of symmetries was given and its properties studied. On the other side, their definition has the shortcoming of being a global one, depending on the structure of the full solution and not on its local physics. The global definition cannot be used, for example, to build an initial condition with a horizon before we solve the evolution. Also, it cannot provide observable physical properties of the UHs, since the global UH does not meet any local conditions there. However, the global UH can be a useful definition in order to analyze global properties of the spacetime, in the same manner that the event horizon definition has been proven to be useful in GR.

In order to numerically study a case of gravitational collapse, a quasilocal definition of UHs has been given in Ref. \cite{Tian:2015vha}, albeit restricted to the case of spherically symmetric spacetimes. That definition consists of replacing the Killing vector field by the Kodama vector field in Eq.~\eqref{staticUH}:
\begin{gather}
K^a u_a = 0 \, , \label{gUHdefinition}
\end{gather}
\noindent
where $K^a$ is the Kodama vector \cite{Kodama:1979vn,Abreu:2010ru}, which can be defined as
\begin{gather}
 K^a = E^{ab}_{\bot} \nabla_b r \, , \label{Kodama-definition}
\end{gather}
\noindent
and $r$ is the areal radius. The definition~\eqref{Kodama-definition} is meaningful only in spherically symmetric spacetimes, because the Kodama vector is only defined in such spacetimes.


Our aim is to give a quasilocal definition of UHs that does not depend on staticity or spherical symmetry. In order to accomplish this, we will translate the definitions we have for symmetric spacetimes in terms of optical scalars associated with a codimension-two foliation $\mathcal{S}$, which are well defined irrespective of the symmetries of the spacetime, provided the assumptions given in Sec.~\ref{sec:contexto} hold.

We start by inserting Eq.~\eqref{Kodama-definition} into Eq.~\eqref{gUHdefinition}, obtaining
\begin{gather}
 0 = K^a u_a = \left(E^{ab}_\perp \nabla_b r \right) u_a  = \left(-E^{ba}_\perp u_a \right) \nabla_b r = - e^b \nabla_b r = 0 \, .
\end{gather}

 The quantity $e^b \nabla_b r$ can be geometrically interpreted. We remark that in spherical symmetry it is proportional to an optical scalar, the 2-expansion $\Theta_{(e)}$:
\begin{gather}
 \Theta_{(e)} = \frac{2}{r} e^a \nabla_a r \, .
\end{gather}
\noindent
 Therefore, we conclude our program stating the general quasilocal definition for UHs as
\begin{gather}
 \Theta_{(e)} = 0 \,. \label{UH-theta-definition}
\end{gather}
 
 The definition \eqref{UH-theta-definition} reduces to the definitions~\eqref{staticUH} and \eqref{gUHdefinition} when the symmetries assumed for those definitions are verified. However, $\Theta_{(e)}$ is well defined even when the leaves of the codimension-two foliation are not spheres, by Eq.~\eqref{theta-definition}, and gives us a view of the UH in terms of the behavior of optical scalars related to the preferred flow $u^a$ across the leaves of $\mathcal{S}$. Therefore, it is a candidate for the \emph{definition of quasilocal UH for general spacetimes}.

 Evidently, different definitions based on optical scalars that also reduce to Eq.~\eqref{UH-theta-definition} in spherically symmetric cases are still possible, if we add terms built with the 2-shear or the 2-vorticity in the $u^a$ or $e^a$ direction, since they vanish in spherical symmetry. However, Eq.~\eqref{UH-theta-definition} already conveys the idea that the future is restricted to 2-surfaces $\mathcal{S}$ of decreasing area and has the advantage of being the simplest general definition based on optical scalars of $\mathcal{S}$ that is equivalent to the current definitions in the particular cases where they apply. We also remark that Eq.~\eqref{UH-theta-definition} is similar in form to the definition of trapping horizons in standard GR\cite{Hayward:1993mw}.
 
 {
 In general spacetimes, there is no equivalent to the areal radius coordinate $r$, and thus we define $\Theta_{e}$ using Eq.~\eqref{theta-definition} which depends only on the foliation $S$ through the induced metric $n_{ab}$. In those cases, there are no natural codimension-two foliations, such as the $S^2$ spheres in symmetrical cases. Hence, we expect that the generalized universal horizons may depend on the choice of the codimension-two foliation, since $e^a$ depends on this choice. This foliation dependence also happens for trapping horizons in general spacetimes and even when using nonspherical foliations in spherically symmetrical spacetimes (see Ref. \cite{Senovilla:2011fk}). 
 }
 
 Once we have a definition for quasilocal UHs in asymmetric spacetimes, we can ask ourselves what is the surface gravity expression on the horizon. Even in static and spherical cases, the surface gravity definition has proved to be a subtle subject, as there are two different prescriptions, 
 \begin{gather}
  \kappa_{\text{inaffinity}} = \left. \sqrt{-\frac{1}{2}\nabla^a \chi^b \nabla_a \chi_b}\right|_{UH} \,, \\
  \kappa_{\text{peeling}} = \left. \frac{1}{2} u^a \nabla_a (\chi^b u_b ) \right|_{UH} \, , \label{eq:kappapeeling}
 \end{gather}
\noindent
whose physical meaning is distinct, as was thoroughly discussed in Ref. \cite{Cropp:2013zxi}. In Ref. \cite{Cropp:2013sea}, it was argued that the surface gravity notion that can be thermodynamically meaningful for modified gravity theories with a UH, is $\kappa_{\text{peeling}}$, since it is the quantity that also appears as related to the temperature of the black hole radiation computed in Ref. \cite{Berglund:2012fk}.

We can write Eq.~\eqref{eq:kappapeeling} in terms of 2-expansions as
\begin{gather}
 \kappa_{\text{peeling}} = \left.- \frac{r}{8} \left( u^a \nabla_a \Theta_{(e)} + \Theta_{u} \Theta_{(e)} \right) \right|_{UH} = \left.- \frac{r}{8} u^a \nabla_a \Theta_{(e)} \right|_{UH} \, . \label{eq:kappatheta}
\end{gather}

We can rewrite it in terms of the horizon area instead of the areal radius $ A_{UH} = 4 \pi r_{UH}^2 = \int_{UH} \mu$ to obtain an expression in terms of quantities that are well defined beyond spherical symmetry:
\begin{gather}
 \kappa_{\text{peeling}} = -\left. \sqrt{\frac{{A_{UH}}}{64 \pi}} \, u^a \nabla_a \Theta_{(e)}\right|_{UH} \, ,\label{theta-kappa}
\end{gather}
\noindent
which shows a remarkable similarity with the \emph{effective surface gravity} defined in Ref. \cite{Senovilla:2014ika} for outer trapping horizons in standard GR,
\begin{gather}
 \kappa_{\text{effective}}= - \left.\sqrt{\frac{{A_{AH}}}{16 \pi}}l^a \nabla_a \Theta_{(k)} \right|_{TH}\,.\label{theta-kappa-ah}
\end{gather}

The use of optical scalars associated with $\mathcal{S}$ and the flow $u^a$ (through $e^a$) allowed us to write the definitions of UH and its surface gravity in the form given in Eqs.~\eqref{UH-theta-definition} and \eqref{theta-kappa}, which can be extended for more general spacetimes. Therefore, we propose Eqs.~\eqref{UH-theta-definition} and \eqref{theta-kappa} as the general definitions of quasilocal UHs and surface gravity, respectively.

It would be interesting to use this surface gravity to find a first-law-of-Thermodynamics type of relation,
\begin{gather}
 \kappa \Lie_z \mu |_{UH} = \Lie_z E |_{UH} + \dots \, ,
\end{gather}
\noindent
where $\mu$ is the area form, $z^a$ is the evolution vector tangent to the UH and $E$ is the quasilocal energy enclosed by the UH. In GR, the Hawking-Hayward energy \cite{Hayward:1993ph} arises as the best candidate. However, the Hawking-Hayward energy is a quantity that depends only on the spacetime metric $g_{ab}$ and seems to be inadequate for theories where the spacetime is also provided with a preferred family of observers given by $u^a$. This is analogous to replacing the trapping horizon notion --- which depends only on $g_{ab}$ --- by the universal horizon, taking into account the role of $u^a$ besides the properties of $g_{ab}$. The definition of a notion of quasilocal energy adapted for nonrelativistic gravity is very interesting in itself; it naturally appears as a result of the application of our formalism to the problem and it is matter for further work.

 \section{Discussion} \label{sec:results}
 
 In this section, we are going to study some immediate consequences using Eq.~\eqref{UH-theta-definition} as the quasilocal UH, and analyze some simple examples.
 
 \subsection{Quasilocal UH and trapped regions}\label{subsec:properties}

 We can rewrite the definition \eqref{UH-theta-definition} in terms of the 2-expansion related to the null directions built with $u^a$ and $e^a$:
 \begin{gather}
  k^a = u^a + e^a \,, \quad l^a = u^a - e^a \, ,
 \end{gather}
\noindent
following the construction made in Ref. \cite{Maciel:2015vva}. In terms of $\Theta_{(k)}$ and $\Theta_{(l)}$, \eqref{UH-theta-definition} takes the form
\begin{gather}
 \Theta_{(k)} - \Theta_{(l)} = 0\, ,\label{UH-definition-null}
\end{gather}
\noindent
which implies $\Theta_{(k)} \Theta_{(l)} > 0$. This means the quasilocal UH  can only appear in the so-called \emph{trapped or antitrapped regions} of spacetime, defined as regions where light rays on the two independent orthogonal directions are either both convergent or both divergent, respectively.

In terms of the known solutions, this only confirms what has been observed in each example: the UH is always contained in the region inside the Killing horizon (or trapping horizon, in the case studied in Ref. \cite{Tian:2015vha}). While trapped regions characterize black holes, antitrapped regions usually appear in expanding spacetimes and white holes. Therefore, we should consider the possibility of a UH of the cosmological type, located in antitrapped regions.

 
 In order to study this possibility, we consider a spatially homogeneous spacetime described by the Friedmann-Lemaître-Robertson-Walker (FLRW) line element
\begin{gather}
 \ud s^2 = - \ud t^2 + a(t)^2 (\ud R^2 + f(R)^2 \ud \Omega^2 ) \, .
\end{gather}
\noindent
where
\begin{gather}
 f(R) = \left\{ \begin{tabular}{l}
                 $R$ --- open flat universe \\
                 $\sinh R $ --- open hyperbolic universe \\
                 $\sin R$ --- closed universe
                \end{tabular} \right.
\end{gather}

The areal radius is given by $r(t,R) = a(t) f(R)$, and its differential is
\begin{gather}
 \ud r = \dot{a}f(R) \ud t + a f' \ud R \, ,
\end{gather}
\noindent
where we use a dot to denote $t$ derivatives and a prime to denote $R$ derivatives.


Irrespective of the full theory, isotropy and spatial homogeneity implies $u^a = \partial_t$ everywhere, {since any component in the $\partial_R$ direction would produce a preferred center of symmetry. For similar reasons, in spherically symmetrical spacetimes, $u^a$ cannot have nonzero angular components}. Thus, we have
\begin{gather}
 e^a = \frac{1}{a} \partial_R  \, .
\end{gather}
\noindent

Therefore, computing $\Theta_{(e)}$, we obtain
\begin{gather}
 \Theta_{(e)} = \frac{2 f'}{a f}\,,
\end{gather}
\noindent
which only vanishes when
\begin{gather}
 f'= 0 \,.\label{eq:condition-cosmo-UH}
\end{gather}

This implies that there are no cosmological UHs in open universes, irrespective of the form of the scale factor. The closed case is more subtle, since Eq.~\eqref{eq:condition-cosmo-UH} is satisfied at the equator ($R = \frac{\pi}{2}$). However, as the closed FLRW spacetime is symmetric about the equator, the region behind it has the same properties as the region in front of it, which means that the equator should not be understood as a quasilocal UH. This happens because this is a surface of extreme radius in each spatial leaf, such that even if the propagation of signals is restricted towards smaller radii, this is not a trapping condition because the radius decreases in both directions, such that physical signals can propagate each way.\footnote{ Something similar happens with the Killing ``horizon'' in the static Einstein Universe, which correspond to our closed example with $a(t) = 1$, and should not be understood as a trapping horizon at all.}

 \subsection{Evolution vector and area}
 
 In order to study the properties of a general UH, it is convenient to define its evolution vector, that is the vector $z^a$ tangent to the UH and restricted to the $(u^a, e^a )$ subspace. Thus, let $z^a =- \gamma u^a + \beta e^a$ be the evolution vector. Here, we need to adopt a criterion in order to choose a direction to $z^a$ and not be ambiguous in what is meant by ``evolution'', as there is an overall sign choice. In the direction tangent to the UH there is no time evolution, as by the UH definition the preferred flow is orthogonal to the UH, and thus we cannot use the prescription usually made for trapping horizons (see Ref. \cite{Hayward:1993mw}) that consists to imposing that evolution vector $z^a$ is future directed.
 
 If we assume spherical symmetry, besides the preferred flow, we have another natural choice which is using the Kodama vector field $K^a$. The Kodama field has the advantage of being proportional to the khronon flow and future directed in the normal region outside the trapping horizon, but it changes continuously when we move inwards until it is orthogonal to the preferred flow at the UH.  The Kodama vector gives us the future direction related to an observer \emph{outside the trapping horizon}. This is the notion of future that we are going to use. Thus, we require that the evolution vector $z^a$ satisfies
 \begin{gather}
  K^a z_a > 0 \,, \label{eq:beta}
 \end{gather}
\noindent which implies $\beta > 0$.
 
 In general, we can use the Hodge dual restricted to $N(\mathcal{S})$ of the mean curvature vector on the leaves of $\mathcal{S}$ instead of the Kodama vector in order to define a vector field $\mathcal{H}^a$ which is reduced to the Kodama vector in spherical symmetry (see this construction in Ref. \cite{Senovilla:2014ika}) and can be used in order to choose a future direction in the same way as done in Eq.~\eqref{eq:beta}.
 
 We have then $z^a z_a = -\gamma^2 + \beta^2$, and 
 \begin{gather}
  \Lie_z \Theta_{(e)}|_{UH} = 0 \Rightarrow -\gamma \Lie_u \Theta_{(e)}|_{UH} +\beta \Lie_e \Theta_{(e)}|_{UH}= 0 \,,\nonumber\\
  \Rightarrow \frac{\gamma}{\beta} =  \left.\frac{\Lie_e \Theta_{(e)}}{\Lie_u \Theta_{(e)}} \right|_{UH} \, .\label{eq:alphabeta}
  \end{gather}
  
  The evolution vector also determines the behavior of the UH area. Consider the area form $\mu = \sqrt{ \det n_{ab} }\epsilon^{\mathcal{S}}_{ab}$, where $\epsilon^{\mathcal{S}}_{ab}$ is the Levi-Civita symbol on $T(\mathcal{S})$. Thus, the evolution of the area measure along the UH is given by
 
 \begin{gather}
  \Lie_z \mu |_{UH} = \mu \Theta_{z} = \mu\left(-\gamma \Theta_{(u)} + \beta \Theta_{(e)}\right)|_{UH} = -\mu \gamma \Theta_{(u)}\,.
 \end{gather}
Since, according to Sec.~\ref{subsec:properties},  the UH lies inside a trapped region,$\Theta_{(u)}<0$. Therefore, the behavior of the area of the horizon depends solely on the sign of $\gamma$:
\begin{gather}
 \left\{ \begin{tabular}{l} 
          $\gamma < 0 \quad  \Rightarrow \quad \Lie_z \mu < 0\, ,$ \\
          $\gamma = 0 \quad  \Rightarrow \quad \Lie_z \mu = 0\, , $\\
          $\gamma > 0 \quad  \Rightarrow \quad \Lie_z \mu > 0\, .$
         \end{tabular} \right.
\end{gather}

  We can then relate the direction of $z^a$ with the (effective) matter sources by means of the evolution equations. Here, for the sake of simplicity, we restrict ourselves to the analysis of the evolution equations in spherically symmetric spacetimes, where only the 2-expansions are nonzero
  \footnote{Those equations can be readily obtained from the evolution equations shown on \cite{Maciel:2015vva}.}:
   
\begin{subequations}\begin{gather}
  \Lie_u \Theta_{(u)} = - \frac{3}{4} \Theta_{(u)}^2 + \frac{1}{4} \Theta_{(e)}^2 + A \, \Theta_{(e)} - \frac{1}{r^2} - 8 \pi T_{ab}e^a e^b \,, \label{LUThetaU} \\
  \Lie_e \Theta_{(e)} = -\frac{3}{4}\Theta_{(e)}^2 + \frac{1}{4} \Theta_{(u)}^2 + B \, \Theta_{(u)}  +  \frac{1}{r^2} - 8 \pi T_{ab}u^a u^b , \label{LEThetaE} \\
  \frac{1}{2}\left(\Lie_u \Theta_{(e)} + \Lie_e \Theta_{(u)} \right) = - \frac{\Theta_{(u)} \Theta_{(e)}}{2} + \frac{A}{2} \, \Theta_{(u)} + \frac{B}{2} \, \Theta_{(e)} - 8 \pi T_{ab}u^ae^b \, , \label{LUThetaE}
  \end{gather}\end{subequations}
\noindent with
\begin{subequations}\begin{gather}
A = u^a e^b \nabla_a u_b \,,\\
B = e^a e^b \nabla_a u_b \,, \label{eq:translation}
 \end{gather}\end{subequations}
 \noindent
 where the physical meanings of $A$ and $B$ are, respectively, the magnitude of the acceleration of $u^a$, $u^a \nabla_a u^b = A e^b$ and $B = K_{ab}e^ae^b$, where $K_{ab}$ is the extrinsic curvature of the leaves of constant time\footnote{In previous articles, such as \cite{Berglund:2012bu}, $A$ and $B$ are denoted as $(a \cdot s)$ and $K_0$, respectively}. We also define the effective energy-momentum tensor $T_{ab}$ as
\begin{gather}
 T_{ab} = T^{\text{matter}}_{ab} + T^{\phi}_{ab}\,,
\end{gather}
\noindent with $T^{\phi}_{ab}$ and $T^{\text{matter}}_{ab}$ representing the energy-momentum tensor related to $S_{\phi}$ and $S_{\text{matter}}$.

Since we can write $\Lie_e \Theta_{(u)} = \Lie_u \Theta_{(e)} - \Theta_{([u,e]}$ and $\Theta_{[u,e]} = A \Theta_{(u)} - B \Theta_{(e)}$, we can rewrite Eq. \eqref{LUThetaE} more conveniently as
\begin{gather}
 \Lie_{u} \Theta_{(e)} = - \frac{\Theta_{(u)} \Theta_{(e)}}{2} + A \, \Theta_{(u)} - 8 \pi T_{ab}u^ae^b \, .
\end{gather}

\subsection{UH formation in truncated HL theory} \label{subsec-collapse}
 
 We analyze the equations of evolution in order to verify conditions in which the UH may be formed at the center $r = 0$ as a result of gravitational collapse in a modified gravity theory. This can be a first step towards an area law analogous to the nondecreasing theorem for trapping horizons (see Ref. \cite{Hayward:1993mw}), such that we have included a version of our reasoning for trapping horizons in the Appendix~\ref{appendix}, which is remarkably simpler.
 
 We consider a truncated nonprojectable HL action in the absence of matter fields (see Ref. \cite{Blas:2010hb}):
 
 \begin{gather}
  S = \frac{1}{16 \pi} \int \ud^4 x \sqrt{-g} \left[ R + (\lambda - 1) (\nabla_a u^a)^2 + \alpha A_{a}A^{a} \right] \, , \label{HL-lagrangian}
 \end{gather}
\noindent
with $A^a = u^b \nabla_b u^a$, where we set $G_{HL} =1$ in order to simplify our notation.

The energy-momentum tensor of the khronon field is given by
\begin{gather}
 8 \pi T^{\phi}_{ab} =  4(1- \lambda) (\nabla_c u^c) \nabla_{(a} u_{b)} - 2 \alpha A_a A_b + g_{ab} \left[ (\lambda - 1) (\nabla_{c} u^{c} )^2 + \alpha A_c A^c \right] \,.
\end{gather}

Specializing for spherical symmetry and projecting in our $(u^a\,, e^a)$ basis, we obtain:
\begin{subequations}\begin{gather}
 8 \pi T^{\phi}_{ab} u^a u^b = (1- \lambda)(B +\Theta_{(u)})^2 - \alpha A^2 \,,\\
 8 \pi T^{\phi}_{ab} e^a e^b =  ( 1 - \lambda)(3 B^2 + 2 B \Theta_{(u)} - \Theta_{(u)}^2 ) - \alpha A^2 \,,\\
 8 \pi T^{\phi}_{ab} u^a e^b = 2 (1 - \lambda ) A (B + \Theta_{(u)}) \, .
\end{gather} \end{subequations}

This lead us to the evolution equations at the UH:

\begin{subequations} \begin{gather}
  \Lie_e \Theta_{(e)} |_{UH} = \left. \frac{1}{4} \Theta_{(u)}^2 + B \, \Theta_{(u)}  +  \frac{1}{r_{UH}^2} -(1- \lambda)(B +\Theta_{(u)})^2  + \alpha A^2 \right|_{UH}\,, \label{LEThetaE-HL} \\
  \Lie_u \Theta_{(e)}|_{UH} = \left. A \, \Theta_{(u)} -2 (1 - \lambda ) A (B + \Theta_{(u)}) \right|_{UH} \, . \label{LUThetaE-HL}
                     \end{gather}\end{subequations}

  We will analyze the general behavior determined by Eqs. \eqref{LEThetaE-HL} and \eqref{LUThetaE-HL} under some assumptions on the value of quantities at the UH that are compatible with a spherical collapse.  First, we consider that $B_{UH} > 0$, which is equivalent to saying that $u^a$ turns inwards as we move towards the center, which happens in all known solutions containing a UH.

  At the UH, we have $u_a = W \nabla_a r$, with $W > 0$, and thus $\Theta_{(u)} = \left.\frac{2}{r_{UH}} u^a \nabla_a r \right|_{UH} = -\frac{2 W^{-1}}{r_{UH}}$. This implies that, if we assume that the spacetime (metric and khronon) is regular until the UH is formed, with $A_{UH}$ and $B_{UH}$  finite when the UH appears at $r = 0$, Eq. \eqref{LEThetaE-HL} is dominated by $\Theta^2_{(u)}$ and curvature terms, both behaving as $\mathcal{O}(1/r^2_{UH})$. 
  
  Hence, we have for $r_{UH} \sim 0$:
  \begin{gather}
   \frac{\gamma}{\beta} = \frac{ \Lie_e \Theta_{(e)}}{\Lie_u \Theta_{(e)}} \sim \frac{\left[1 -W^{-2}(3-4\lambda)\right] (1/r^2_{UH}) }{A \Theta_{(u)} (2\lambda-1)   }
  \end{gather}

  In addition, we consider $A_{UH} > 0$, which happens in the analytic solutions found in Ref. \cite{Berglund:2012bu} and means that the fluid flow does not turn inwards as fast as the geodesic flow.
  
  Hence, we have that the denominator is negative for $\lambda > 1/2$. In order to analyze the sign of the numerator we remark that at the UH we have $-1 = u^a u_a = W^2 \nabla^a r \nabla_a r \Rightarrow W^{-2} \sim \frac{2 M_{MS}(r = 0)}{r_{UH}} -1$, where $M_{MS}$ denotes the Misner-Sharp energy. If $\lim_{r \to 0} M_{MS} > 0$, this term dominates the numerator, and we have that the numerator is negative for $\lambda < 3/4$. Therefore, under the assumptions above, we only have $\gamma > 0$ for $1/2 < \lambda < 3/4$. In this case, a UH  formed can have an increasing area. Otherwise, at $r=0$, the derivative of the area of a UH formed at the center would be negative ($\gamma < 0$) which implies that the UH would instantly vanish. 
  
  Recapitulating, under the assumptions that
  \begin{itemize}
   \item gravity is described by the truncated HL Lagrangian in Eq.~\eqref{HL-lagrangian},
   \item for $r \to 0$, $A> 0$ and $B > 0$, but finite, and
   \item for $r \to 0$, the Misner-Sharp energy $M_{MS} (r \to 0) > 0$,
  \end{itemize}
  \noindent
  the smooth formation of a UH increasing from the center $(r = 0)$ is only allowed for  $1/2 < \lambda < 3/4$.
  
  If we consider $A_{UH}< 0$ instead, keeping the other assumptions, we obtain the increasing UH area for $\lambda < 1/2$ or $\lambda > 3/4$.
  
  For other values of $\lambda$, under our assumptions, our reasoning does not exclude the possibility of UH formation, but it implies that the UH can only appear at finite radius, resulting in the sudden formation of a black hole with a finite area. This can be interesting from the point of view of the thermodynamical interpretation of the laws of black holes dynamics in nonrelativistic gravity theories, since the thermodynamical variables related to the black hole horizon area should display a discontinuous behavior in those cases.
  
  Reference \cite{Tian:2015vha} showed a numerical simulation of a spherical gravitational collapse in Einstein-\ae ther theory, but the beginning of UH formation was not ploted and in all continuous evolution, the area of the UH decreases, as can be readily checked in their Fig. 2. An extension of this kind of work in order to analyze the event where the UH is formed would be welcome in order to better understand our results above.

%
%

 \section{Conclusion}\label{sec:conclusions}
 
 In this paper we reviewed of the universal horizons definitions in the literature --- which are well defined only under strong symmetry assumptions --- and translated it into the language of optical scalars, namely, 2-expansions of the flows related to the preferred foliation of the spacetime. We also have translated the surface gravity definition associated with the peeling behavior of trajectories near the universal horizon. With these translated definition we eliminated the dependence on specific spacetime symmetries and obtained a robust definition under much weaker spacetime assumptions --- as explained in Sec.~\ref{sec:contexto} --- while keeping the quasilocal aspect needed in order to deal with dynamical situations. 
 
 Using our new formalism, we have shown that quasilocal universal horizons are always restricted to trapped or antitrapped regions of the spacetime. Then, we dealt with the possibility of the existence of cosmological universal horizons --- corresponding to a horizon in an antitrapped region ---  and proved they do not exist in FLRW models, irrespective of the scale function $a(t)$. This result supports the notion that the universal horizons only appear in black holes.
 
Finally, we analyzed the properties of the evolution equations of the quasilocal universal horizons in the spherically symmetric case. By making simplifying assumptions we were able to show that in the case of the collapse governed by a truncated Ho\v rava-Lifshitz theory, there are cases in the parameter space in which the universal horizons cannot be formed starting smoothly from the center, contrary to what intuitively seems more natural. This can have interesting implications for the corresponding black hole thermodynamics, since a discontinuous formation should have a thermodynamical counterpart.
 
 
 There is another general definition of universal horizons in the literature (given by Bhattacharyya \emph{et al.}) that also has the advantage of not relying on spacetime symmetries, but it is a global definition, that depends on the full causal structure of spacetimes with a preferred foliation, whose theory has been described with great precision in Ref. \cite{Bhattacharyya:2015gwa}. The relationship between our quasilocal definition and the global definition of universal horizons should be investigated. Is it analogous to the relation between trapping horizons and event horizons? It would be interesting to study the formation and behavior of the two kinds of horizons in a numerical simulation of a gravitational collapse, for example.
 
Another interesting issue is the analysis of the evolution equations~\eqref{LUThetaU}, \eqref{LEThetaE} and \eqref{LUThetaE} --- and their generalization beyond spherical symmetry --- in order to obtain formulas analogous to the laws of black dynamics given for standard GR black holes in Ref. \cite{Hayward:1993mw}. Such laws (if they exist), due to their thermodynamical flavor, are steps toward a consistency test for the underlying modified gravity theories through thermodynamics.

Given the similarities between the expressions found in this work for quasilocal universal horizons in our formalism and the analogous expressions related to trapping horizons, this approach  not only opens the way to the study of the general gravitational collapse in nonrelativistic gravity; it also suggests that we should repeat the trapping horizon program that has been fruitful in giving a better understanding of extreme regimes in relativistic gravity and may also give us a new way to look at theories of gravity with a preferred foliation.

\begin{acknowledgments}
 I thank D. Guariento, V. Zanchin, C. Molina, M. Le Delliou, M. Mello and M. Rodrigues for helpful discussions. This work was supported by CAPES, Grant n\textordmasculine~88881.064999/2014-01. 
\end{acknowledgments}

\appendix

\section{Trapping horizon formation in GR}\label{appendix}

Here, we show that trapping horizons (THs) that appear at the center are nondecreasing. Actually, this is just a particular case of the way more general nondecreasing area theorem by Hayward \cite{Hayward:1993mw}, which we are using to illustrate the reasoning the lead us to the results of Sec.~\ref{subsec-collapse}.

We assume that
\begin{itemize}
 \item $T_{ab}$ respects the null energy condition: $T_{ab}K^a K^b \geq 0$ for any null $K^a$, and
 \item $T_{ab}$ is regular at the center until the formation of the TH.
 \end{itemize}

Therefore, using a null basis built with our $(u,e)$ basis as
\begin{gather}
 k^a = u^a + e^a \,,\\
 l^a = u^a - e^a \,,
\end{gather}
\noindent
a black hole TH is defined as the hypertube foliated by surfaces where
\begin{gather}
 \Theta_{(k)} = 0 \, , \quad
 \Theta_{(l)} < 0\,,
\end{gather}
\noindent
in the black hole case. The analogy between the pairs $k^a/e^a$ and $l^a/u^a$ is clear. The evolution vector $z^a = -\gamma l^a + \beta k^a$, with $\beta > 0$, satisfies:
\begin{gather}
0= \Lie_z \Theta_{(k)} |_{TH} = -\gamma \Lie_l \Theta_{(k)} + \beta \Lie_k \Theta_{(k)} \Rightarrow \frac{\Lie_k \Theta_{(k)}}{\Lie_l \Theta_{(k)}} = \frac{\gamma}{\beta}\,.
\end{gather}

We use the evolution vector to compute the area evolution:
\begin{gather}
 \Lie_z \mu|_{TH} = \mu \Theta_{(z)} |_{TH} = -\mu \gamma \Theta_{(l)}\,.
\end{gather}

Thus, $\gamma > 0$ implies that the TH area is increasing.

Let us study the behavior of a TH at the center $r \to 0$, using the evolution equations:
\begin{gather}
 \Lie_k \Theta_{(k)} = - \frac{ \Theta_{(k)}^2 }{2} + \nu_k \Theta_{(k)} - 8 \pi T_{ab} k^a k^b \, ,\\
 \Lie_l \Theta_{(k)} = \frac{1}{2} \Theta_{([l,k])} - \Theta_{(k)}\Theta_{(l)} - \frac{2}{r^2} + 8 \pi T_{ab}l^a k^b \,.
 \end{gather}

 At the TH, $\Theta_{(k)}$ vanishes and we have
 \begin{gather}
  \Lie_k \Theta_{(k)}|_{TH}= - 8 \pi T_{ab}k^a k^b \leq 0 \, , \\
  \Lie_l \Theta_{(k)}|_{TH}=  \frac{1}{2} \Theta_{([l,k])} - \frac{2}{r^2} + 8 \pi T_{ab}{l^a k^b} \sim - \frac{2}{r^2} + \mathcal{O}(r^{-1}) < 0 \, ,
 \end{gather}
 \noindent
 where the first inequality comes from the null energy condition and the second comes from the regularity of $T_{ab}$ at the center. Hence, for a TH with $r \to 0$, we have, under the above conditions,
\begin{gather}
 \frac{\Lie_k \Theta_{(k)}}{\Lie_l \Theta_{(k)}} \geq 0 \Rightarrow \gamma \geq  0,
\end{gather}
\noindent
which implies that any trapping horizon forming at the center has a nondecreasing area, which is the expected behavior.

\bibliography{shortnames,referencias}
\end{document}